*Article*

# Developing a Climate Litigation Framework: China's Contribution to International Environmental Law

Yedong Zhang *

Law School of Fudan University, PRC No.220 Gengdan Road, Shanghai 200438, China

* Corresponding author. E-mail: 21110270022@m.fudan.edu.cn (Y.Z.)

**ABSTRACT:** Although "climate litigation" is not an indigenous term in China, localizing it is essential to support the development of an independent environmental legal knowledge system in China. Rooted in China's judicial tradition, which emphasizes substantive rationality, traditional legal theories have primarily focused on environmental law. However, the contemporary practices in the rule of law have created an unclear trajectory for climate litigation. Research in this area has long been trapped in a paradigm that relies on lawsuits for ecological environmental damage compensation and environmental public interest litigation, leading to a significant disconnect between theoretical frameworks and practical application. With the advancement of the "dual carbon" strategic goals—carbon peaking and carbon neutrality—it has become imperative to redefine the concept of climate litigation within the Chinese context. We need to establish a theoretical framework that aligns with the "dual carbon" objectives while providing theoretical and institutional support for climate litigation, ultimately contributing to the international discourse on climate justice. Additionally, Hong Kong's proactive climate governance and robust ESG (Environmental, Social, and Governance) practices provide valuable insights for developing comprehensive climate litigation mechanisms. Based on this analysis, we propose concrete plans for building a climate litigation system in China, establishing a preventive relief system and a multi-source legal framework at the substantive level and developing climate judicial mechanisms for mitigation and adaptation at the procedural level.

**Keywords:** Climate litigation; Dual carbon goals; ESG practices; International environmental law; Climate governance

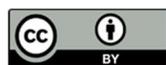

## 1. Introduction

Climate litigation is a powerful tool to enforce and drive policy shifts toward cleaner energy within the framework of climate law [1]. This article examines climate lawsuits as a distinct mechanism within the broader frameworks of climate policy and climate law. While climate policy and law set the overarching framework for addressing climate change, climate lawsuits are a targeted tool for enforcing, advancing, or challenging these policies [2]. By focusing on cases brought before courts to address issues of climate accountability, responsibility, and regulatory compliance, this paper highlights the distinctive role of litigation in driving climate action. The analysis underscores how these lawsuits intersect with broader legal and policy considerations yet remain distinct as strategic legal interventions [3]. The scope of climate litigation encompasses a broad range of legal actions aimed at addressing the impacts of climate change through judicial means. This field includes lawsuits filed to enforce regulations and policies on greenhouse gas emissions, protect human rights affected by climate change, and seek compensation for climate-related damages. Climate litigation serves to hold governments and corporations accountable for their environmental responsibilities, helping to ensure compliance with climate commitments and regulations. It addresses both the mitigation of climate change by reducing emissions and emphasizes adaptation strategies to manage the effects of climate disruptions on various sectors of society. This dual focus is essential for a holistic approach to combating climate change, as it addresses both the reduction of future risks through mitigation but also by enhancing resilience through adaptation. This paper conducts empirical research on climate litigation by analyzing various types of cases and their outcomes within different legal frameworks. The study draws on a comprehensive dataset of climate litigation cases from multiple jurisdictions, including landmark cases from the United States and the European Union and emerging examples from countries like

Pakistan and the Philippines. The empirical study focuses on understanding the effectiveness of climate litigation in achieving environmental justice and promoting regulatory compliance [4]. By examining real-world case data, this research aims to identify patterns, challenges, and best practices in climate litigation, providing insights that can inform future legal strategies and policy-making [5]. The analysis reveals key trends such as the increasing use of human rights arguments in climate cases and the strategic litigation aimed at pushing for more ambitious climate policies. It also highlights the diverse approaches taken by different courts in addressing climate issues, reflecting varying levels of judicial activism and interpretation of environmental laws. By situating China's climate litigation within this international context, the paper aims to contribute to the global discourse on climate justice while proposing a unique framework tailored to China's specific legal and environmental conditions. This includes exploring how international legal principles can be integrated with domestic law, thereby enhancing the effectiveness of climate litigation in achieving both national and global environmental goals.

This article defines "climate lawsuits" broadly, encompassing legal actions aimed at advancing climate goals rather than limiting the scope to lawsuits that challenge legislative bodies for stronger climate laws based on human rights claims in a liberal-democratic context. Given China's distinct legal framework, which fundamentally differs from systems that emphasis judicial checks and balances, climate litigation in China focuses on integrating environmental objectives within existing legal and policy structures. This research aims to explore how climate litigation, though rooted in a unique system, can still contribute to enforcing environmental responsibilities and advancing China's "dual carbon" goals within its legal context. This article also considers climate lawsuits targeting specific facilities or companies responsible for significant emissions, such as factories or airports, within China's regulatory and industrial context. Such cases pose unique challenges: enforcing climate targets at an individual facility level requires navigating complex local licensing laws and confronting the practical issue of allocating national emissions targets to single sources. Additionally, liability lawsuits against companies raise attribution challenges, as linking specific climate impacts to individual emitters remains difficult without advanced attribution science. This study acknowledges these limitations, exploring how China might address them within its legal and environmental frameworks. The role of courts and litigants in shaping climate governance through judicial action [6], climate litigation is advancing legal and regulatory frameworks to address climate change as a justice issue [7]. Climate change disrupts traditional legal frameworks, necessitating innovative regulatory responses [8]. Global trends in climate litigation demonstrate the increasing role of the judiciary in enforcing climate commitments and protecting human rights. Landmark cases such as Urgenda Foundation v. State of the Netherlands [9] and Leghari v. Federation of Pakistan [10] highlight the judiciary's role in mandating governmental adherence to climate policies [11]. The integration of human rights into climate litigation, as seen in Greenpeace Southeast Asia v. Carbon Majors [12], underscores the duty of states and corporations to protect the right to a healthy environment. Furthermore, innovative legal approaches, such as those employed in Ralph Lauren 57 v. Byron Shire Council in Australia [13], emphasize the importance of adaptation in legal strategies. The Paris Agreement provides a crucial framework for these actions, enabling litigants to challenge insufficient national policies based on international commitments [14]. By situating China's climate litigation within this international context, this research aims to contribute to the global discourse on climate justice while proposing a unique framework tailored to China's specific legal and environmental conditions. In addition to these global examples, Hong Kong has made significant strides in climate governance through its robust ESG (Environmental, Social, and Governance) practices and proactive climate policies [15]. The Hong Kong Climate Action Plan 2030+ outlines the city's strategic approach to reducing carbon emissions and enhancing climate resilience, while the mandatory ESG reporting requirements by the Hong Kong Stock Exchange (HKEX) set a high standard for corporate transparency and accountability in environmental matters [16]. These initiatives contribute to local sustainability efforts and offer valuable lessons for integrating climate litigation with broader governance frameworks. By situating China's climate litigation within this international context and drawing on Hong Kong's advanced ESG practices, this research aims to contribute to the global discourse on climate justice while proposing a unique framework tailored to China's specific legal and environmental conditions. This comprehensive approach seeks to bridge the gap between theory and practice, providing practical solutions to enhance the effectiveness of climate litigation in achieving both national and international environmental goals.

## 2. Integrated Foundation: The Jurisprudential Basis of the Climate Litigation

To build a climate litigation system that aligns with international standards and addresses global concerns, it is essential to analyze the legal structure of climate litigation, focusing on the system's rights and obligations. From a rights perspective, compared to developed nations like the United States and those in Europe, China's climate litigation practice is emerging, influenced by its unique legal and political context. In these established jurisdictions, climate litigation has a longer history, supported by well-documented cases that have shaped legal precedents and regulatory frameworks. For example, cases like Massachusetts v. EPA in the United States and the Urgenda case in the Netherlands have significantly influenced climate policies and enforcement mechanisms. China's climate litigation practice, on the other hand, is still in a developmental phase. This emerging practice is characterized by its integration into the broader framework of environmental public interest litigation [17], a mechanism that has been primarily used for addressing ecological damage and pollution. Given the global emphasis on international cooperation and environmental governance, China's engagement in international carbon governance highlights the importance of bridging the gap between theory and practice by examining liability attribution [18], damage assessment, and causality in climate litigation. The attribution of liability in climate cases often involves complex scientific evidence and causation theories, which can be challenging to establish in court. Damage assessment in climate litigation also requires a sophisticated understanding of environmental science and economics to quantify the impacts of climate change accurately. Furthermore, establishing causality links the defendant's actions to the harm caused, necessitating robust legal and scientific frameworks to support these claims. By aligning its climate litigation system with international standards, China can enhance its legal framework to address these challenges more effectively. This alignment involves adopting best practices from other jurisdictions, integrating international legal principles, and developing comprehensive laws that support climate litigation. This approach will strengthen China's domestic climate governance and contribute to the global effort to combat climate change through effective legal mechanisms.

*2.1. The Rights Basis of Climate Litigation*

Analyzing the rights basis of climate litigation requires an empirical approach to clarify the current challenges. The United States offers valuable insights with its extensive history of climate litigation. Scholars typically classify American climate litigation into cases targeting government agencies and private entities. According to the International Bar Association, these cases often rely on statutes such as the Clean Air Act and the National Environmental Policy Act, which provide a legal foundation for addressing climate-related harms. The Sabin Center for Climate Change Law documents numerous cases where plaintiffs have sought judicial redress for climate impacts, leveraging these statutory frameworks [19]. One prominent example is Connecticut v. American Electric Power Co., where plaintiffs used public nuisance to address climate-related tortious conduct [20]. Despite judicial challenges, there is a growing trend toward awarding damages or injunctions in such cases, reflecting an evolving judicial recognition of the need to tackle climate issues through legal means. Reports by the United Nations Environment Programme and the Sabin Center highlight the increasing engagement of courts with the complex scientific and legal questions posed by climate litigation. This trend is further supported by strategic human rights arguments and the incorporation of international environmental norms [21]. Additionally, U.S. climate litigation often includes constitutional challenges, invoking provisions of the Commerce Clause and the First Amendment to advocate for stronger climate policies. This multifaceted approach underscores the intersection of environmental law with broader legal principles. The empirical study of these cases provides insights into effective legal strategies for achieving environmental justice and regulatory compliance [22], ultimately contributing to the development of robust legal frameworks for addressing climate change globally [23].

Furthermore, compared to other countries, a higher proportion of U.S. climate litigation involves government agencies as defendants. Based on the data in Table 1, it is evident that the National Environmental Policy Act claims account for the highest number of federal statutory climate litigation cases in the United States, with a total of 348 cases, followed by claims under the Endangered Species Act and other wildlife protection laws, totaling 194 cases. Environmental organizations have leveraged the public trust doctrine to sue governments and administrative agencies, compelling them to formulate or enforce stricter climate policies to mitigate the severe impacts of climate change. This legal principle asserts that the government must protect natural resources for public use, serving as a powerful tool in environmental advocacy. In the common law system, various types of climate litigation have gradually established precedent, becoming integral to U.S. environmental law and policy. Here, "precedent" refers not only to prior cases with similar facts but also to a broader accumulation of decisions that collectively shape legal standards and influence future cases in the realm of environmental regulation. For instance, the case of Juliana v. United States, where plaintiffs argued that the government's inaction on climate change violated their constitutional rights, has significantly influenced the legal discourse on governmental responsibility. Although facing procedural challenges, it set a foundation for future litigation by highlighting the judiciary's role in addressing climate issues. The strategic use of the public trust doctrine and other legal principles has enabled environmental organizations to hold governments accountable. This trend is evident in cases like Massachusetts v. EPA, where the U.S. Supreme Court recognized the EPA's authority to regulate greenhouse gases under the Clean Air Act. Such cases underscore the judiciary's role in enforcing environmental regulations and shaping climate policy [24]. Reports by international bodies, such as the Parliamentary Assembly of the Council of Europe, emphasize the importance of legal frameworks in protecting the right to a healthy environment. These reports advocate for integrating environmental rights into human rights frameworks, reflecting a global trend towards recognizing and enforcing environmental protections through judicial means [25].

Table 1. Types and Data of Climate Litigation Cases in the United States.

| | Types of Claims | Number of Cases |
|---|---|---|
| Federal Statutory Claims | Clean Air Act | 193 |
| | Endangered Species Act and Other Wildlife Protection Laws | 194 |
| | Clean Water Act | 63 |
| | National Environmental Policy Act | 348 |
| | Freedom of Information Act | 87 |
| | Other Statutory Claims | 189 |
| Constitutional Claims | Commerce Clause | 23 |
| | First Amendment | 12 |
| | Fifth Amendment | 16 |
| | Fourteenth Amendment | 19 |
| | Other Constitutional Claims | 41 |
| State Law Claims | Freedom of Information/Public Records | 24 |
| | Other Types of State Law Cases | 29 |
| | Industry Litigation | 55 |
| | Environmental Litigation | 76 |
| | State Impact Assessment Laws | 257 |
| | Enforcement Cases | 17 |
| | Public Utility Regulation | 53 |
| | Common Law Claims | 32 |
| | Public Trust Claims | 28 |
| | Securities and Financial Regulation | 29 |
| | Trade Agreements | 1 |
| Adaptation | Other Types of Adaptation Cases | 2 |
| | Reverse Impact Assessments | 27 |
| | Lawsuits Seeking Monetary Compensation for Losses | 33 |
| | Insurance Cases | 5 |
| | Challenges to Adaptation Measures | 24 |
| | Actions Seeking Adaptation Measures | 39 |
| Protesters and Scientists | Protesters | 41 |
| | Scientists | 15 |

China's climate litigation research is particularly important, given its late start compared to developed nations like the United States, as it lays the groundwork for understanding how legal mechanisms can be utilized to address environmental issues. Moreover, the limited number of practical judicial cases underscores the need for comprehensive research to inform and shape future litigation strategies. Following the signing of the Paris Agreement and the establishment of the "dual carbon" goals, climate change response was integrated into the framework of national governance, and climate litigation was officially incorporated into the scope of environmental public interest litigation in China. This marked a significant shift in the country's approach to environmental law. One notable case is the "Friends of Nature" litigation against State Grid Ningxia Electric Power Company and State Grid Gansu Electric Power Company over "abandoning wind and solar" practices. These cases represent the beginning of climate litigation in China, setting important precedents for future legal actions. They demonstrate the judiciary's increasing role in addressing environmental issues and enforcing climate policies. Despite these advancements, challenges remain. The legal framework for climate litigation is still developing, and there is a need for more comprehensive laws specifically addressing climate issues. The judiciary also requires further specialization and training to handle the scientific and technical aspects of these cases effectively. However, significant opportunities exist. The integration of climate goals into national policies provides a strong foundation for future litigation. Growing public awareness and international focus on climate change can also bolster these efforts. By learning from global experiences and adapting them to local contexts, China can enhance its legal responses to climate change. Overall, while still in its early stages, China's progress in climate litigation shows a strong commitment to addressing climate change through judicial means, promising substantial advancements in environmental protection.

Despite China's late start and limited practice in climate litigation, with theoretical research lagging behind judicial practice, the lack of a supporting legal framework has left judicial authorities ill-prepared to handle climate-related cases effectively. As a result, the judiciary often struggles to fulfill its role in resolving disputes efficiently. Bridging the gap between theory and practice in China's climate litigation is significant [26]. This involves addressing the challenges posed by the evolving nature of climate law and the necessity for a more robust legal infrastructure to support climate litigation efforts. The theoretical underpinnings of climate litigation in China are still developing, and this gap between theory and judicial practice needs urgent attention. It is crucial to promptly analyze and integrate the various issues encountered in judicial practice to bridge this gap effectively. A key area requiring attention is the establishment of a comprehensive legal framework that encompasses both substantive and procedural laws specifically tailored to climate-related disputes. This framework should draw on international experiences while adapting to China's unique legal and environmental context. The rights basis of climate litigation is fundamental to understanding and advancing climate litigation. It often derives from fundamental human rights, such as the right to life, health, and development. These rights are increasingly recognized as being threatened by climate change, thus forming the basis for legal actions to hold governments and corporations accountable for their contributions to climate-related harm. Integrating human rights into climate litigation provides a powerful tool for advocates seeking to compel action on climate issues. Moreover, the development of specialized legal principles and doctrines tailored to climate litigation is essential. This includes the incorporation of principles like the precautionary principle and the polluter pays principle, which can guide judicial decision-making in climate cases. Additionally, enhancing the judiciary's capacity through training and the establishment of specialized environmental courts can improve the effectiveness of climate litigation. China's engagement in international climate governance and its commitment to the "dual carbon" goals provide a strong impetus for developing a robust climate litigation framework. By learning from global best practices and incorporating them into local judicial practices, China can create a climate litigation system that not only addresses domestic environmental challenges but also contributes to the global effort to combat climate change. Overall, the establishment of a well-supported legal framework for climate litigation in China, informed by both theoretical research and practical judicial experiences, is crucial for bridging the gap between theory and practice. This will enable the judiciary to play a more effective role in resolving climate-related disputes and advancing environmental justice.

*2.2. The Obligations Basis of Climate Litigation*

In the realm of climate litigation, the basis of the obligation is critical and includes two aspects: the state's duty of care and the obligation of scientific provision, the latter of which should be defined here or placed in quotation marks to indicate that it is a term of art that will be explained later. As carbon emitters, entities such as corporations, industries, and other organizations are responsible for exercising a duty of care. If carbon emitters breach this duty, they can be held liable for climate change-related torts. The state's duty of care is foundational in the context of climate change, requiring active engagement in addressing climate change, recognizing its impacts on human society and the natural environment, and taking necessary measures to prevent or mitigate these impacts. This duty involves not only protecting specific individuals but also defending society from broader, abstract threats posed by climate change [27]. The duty of care requires the government to take proactive steps to safeguard the environment and protect public health. Formulating and enforcing stringent regulations on greenhouse gas emissions, implementing policies to reduce carbon footprints, and ensuring compliance with international climate agreements. Failure to fulfill these obligations can result in legal actions, holding the state accountable for climate-related damages. Moreover, the obligation of scientific provision requires the state to base its climate policies and actions on robust scientific evidence. This means that the state must support and utilize scientific research to understand the causes and effects of climate change, develop effective mitigation and adaptation strategies, and assess the potential impacts of various policies and projects. By grounding climate actions in scientific knowledge, the state can ensure that its efforts are effective and credible, thereby fulfilling its legal and moral obligations to protect the environment and public health [28]. The Supreme People's Court of the People's Republic of China emphasizes the importance of this dual responsibility in its environmental adjudication guidelines. These guidelines highlight the need for a legal framework that integrates scientific principles and ensures that the state and other entities uphold their duty of care in climate-related matters. By doing so, the judiciary can play a pivotal role in advancing climate justice and promoting sustainable development [29].

The obligation of scientific provision requires the state to act based on scientific knowledge and principles in formulating and implementing climate change policies. The state's climate change response measures are grounded in scientific research to ensure policy effectiveness and scientific validity. This obligation ensures that climate policies are not merely reactive but are strategically designed to effectively address the complex and evolving nature of climate change [30]. To meet this obligation, the state must include national target clauses in legal texts that support and promote scientific research on the causes, impacts, and strategies for climate change adaptation and mitigation. These national target clauses act as benchmarks for measuring progress and accountability, ensuring that the state's climate actions align with the latest scientific findings and international standards. This integration of scientific research into policy-making is critical for developing robust climate strategies that can withstand legal scrutiny and achieve tangible environmental benefits. In essence, the obligation of scientific provision in climate litigation stems from the state's environmental protection duty established in the Constitution of the People's Republic of China. This constitutional duty mandates that the state protect the environment and ensure sustainable development, thereby providing a legal basis for incorporating scientific principles into climate governance. By grounding climate policies in scientific evidence, the state can enhance the legitimacy and effectiveness of its climate actions, fostering public trust and international cooperation [31]. Moreover, this obligation underscores the importance of transnational climate litigation, as discussed in the literature on the contribution of the Global South to climate justice. Transnational litigation can play a crucial role in holding states accountable for their climate actions, promoting the dissemination of scientific knowledge across borders, and encouraging global collaboration in addressing climate change [32]. The case of Ali v. Federation of Pakistan, where the court emphasized the need for scientifically informed policies to address climate change effectively. Similarly, scholarly discussions, such as those by Jamieson on climate responsibility and justice, highlight the ethical imperatives for states to base their climate actions on sound scientific principles. By integrating scientific research into legal and policy frameworks, the state not only fulfills its constitutional and ethical duties but also sets a precedent for proactive and informed climate governance. This approach ensures that climate litigation can effectively address the complexities of climate change, promoting resilience and sustainability in the face of global environmental challenges [33].

Overall, the rights and obligations basis of climate litigation must be unified and directed toward corresponding climate responsibilities, forming the jurisprudential foundation of the climate litigation system. This unity is found in the right to climate stability under environmental rights and the state's duty of care under the obligation of scientific provision [34]. In climate litigation, the state's failure to fulfill its climate obligations directly affects the realization of climate rights. Thus, the state's inaction or wrongful actions could constitute an infringement of these rights. Climate litigation should address current rights violations and prevent future damage [35]. In the context of China's climate litigation practice, although a corresponding climate litigation system has not yet been fully established, the state's environmental protection obligations have gained recognition.

## 3. Two Dimensions: Analytical Dimensions of the Climate Litigation System

The climate litigation system can be categorized into two types: mitigation-focused and adaptation-focused, with each type examined through the lenses of their respective goals and strategies. The distinction between mitigation and adaptation in climate litigation is derived from Article 4, Section 1(f) of the United Nations Framework Convention on Climate Change (UNFCCC): "All Parties, taking into account their common but differentiated responsibilities and their specific national and regional development priorities, objectives, and circumstances, shall: (f) Take climate change considerations into account, to the extent feasible, in their relevant social, economic, and environmental policies and actions, and employ appropriate methods, for example, impact assessments, formulated and determined nationally, with a view to minimizing adverse effects on the economy, public health, and the quality of the environment, of projects or measures undertaken by them to mitigate or adapt to climate change". This provision emphasizes that climate action should address mitigation and adaptation dimensions, though the UN language suggests these can be considered alternative approaches. Mitigation-focused climate litigation involves legal actions to reduce greenhouse gas emissions and holding parties accountable for contributing to climate change. This includes cases against corporations for excessive emissions, challenges to inadequate government policies, and enforcement of existing environmental regulations. Examples include lawsuits filed to enforce compliance with emission reduction targets; however, Massachusetts v. EPA was focused on the EPA's authority to regulate greenhouse gases under the Clean Air Act rather than enforcing specific emission reduction targets. Supreme Court recognized the authority of the Environmental Protection Agency to regulate greenhouse gases under the Clean Air Act. Adaptation-focused climate litigation, on the other hand, deals with legal actions that aim to manage and adapt to the impacts of climate change. This includes lawsuits seeking to compel governments and corporations to take measures to protect communities from the adverse effects of climate change, such as rising sea levels, extreme weather events, and disruptions to food and water supplies. A notable example is the Leghari v. Federation of Pakistan case, where the court-mandated the government to implement its climate adaptation policies to protect the rights of citizens affected by climate change. The unity of climate rights and obligations is reflected in the concept of climate responsibility, which includes not only mitigating climate change impacts but also adapting to and preventing future potential damage. This dual approach establishes a comprehensive legal framework that addresses both the causes and impacts of climate change, fostering resilience and sustainability. By integrating both mitigation and adaptation strategies, the climate litigation system can more effectively tackle the multifaceted challenges posed by climate change. This integrated approach is essential for fostering a legal environment that supports sustainable development and environmental justice on both national and international levels.

*3.1. Mitigation Dimension*

Climate mitigation refers to reducing greenhouse gas emissions from human activities. Key measures include increasing forestry's capacity to absorb greenhouse gases, reducing high-emission energy consumption, and improving energy efficiency. These actions are foundational to the theoretical basis for the "dual carbon" goals, which aim for carbon peaking and neutrality. From the perspective of social systems theory, achieving these goals requires translating policy discourse into concrete, enforceable actions, and it is essential to explore how to achieve coherent and orderly normative translation within the legal system, ensuring that climate policies are not only well-defined but also effectively implemented and enforced. Mitigation technologies must be developed and mobilized while avoiding regulatory deadlocks between businesses and regulators. Fostering innovation in clean energy technologies, enhancing energy efficiency standards, and promoting the use of renewable energy sources. Legal frameworks must support these technological advancements by incentivising green technology adoption and imposing penalties for non-compliance with emission reduction targets .

From the perspective of interest distribution, achieving the "dual carbon" goals affects the distribution of interests across various sectors, necessitating legal measures to manage conflicts and ensure coordination. A sound legal system, strict enforcement, impartial justice, and effective social oversight. Legal mechanisms must be in place to address disputes that arise from the transition to a low-carbon economy, ensuring that the interests of all stakeholders are considered and balanced. The primary dimension in constructing a climate litigation system is mitigation, specifically mitigation-focused climate litigation. This type of litigation often requires scientific evidence to establish a causal link between certain actions or policies and climate change [36]. Litigants may need to demonstrate how specific emissions contribute to global warming and its associated impacts. This process involves comprehensive data collection, advanced modeling techniques, and expert testimony to substantiate claims [37]. Mitigation-focused climate litigation may involve upgrading traditional industries to reduce their carbon footprint, promoting the adoption of clean energy technologies, and supporting emerging sectors with electric vehicles. These legal actions can compel industries to adhere to stricter emission standards, accelerate the deployment of renewable energy projects, and foster the development of sustainable transportation options. By holding emitters accountable and enforcing compliance with environmental regulations, mitigation-focused litigation can drive significant reductions in greenhouse gas emissions and contribute to global climate goals.

Mitigation litigation also requires public participation, which can be achieved by incorporating the environmental public interest litigation system and ensuring public involvement in climate governance through information disclosure and public hearings. Public participation is crucial because it enhances transparency, accountability, and the democratic legitimacy of climate policies and actions. Engaging the public in the litigation process helps to ensure that the diverse interests and concerns of different stakeholders are considered, fostering a more inclusive and comprehensive approach to climate mitigation. Mitigation-focused climate litigation can be divided into three categories: (1) administrative lawsuits related to greenhouse gas allocation and confirmation, (2) administrative public interest lawsuits for insufficient greenhouse gas reduction by administrative agencies, and (3) civil public interest lawsuits for corporate climate information disclosure and illegal greenhouse gas emissions. Each category addresses different aspects of climate mitigation and involves distinct legal principles and procedural rules. Administrative lawsuits related to greenhouse gas allocation and confirmation typically involve challenges to government decisions on the distribution and regulation of emission permits. These cases often focus on whether the allocation aligns with national and international emission reduction targets and whether the process is fair and transparent. Administrative public interest lawsuits for insufficient greenhouse gas reduction by administrative agencies involve claims that government bodies have failed to implement or enforce policies that adequately address climate change. These lawsuits compel agencies to take stronger action to meet their legal obligations and protect the environment. Civil public interest lawsuits for corporate climate information disclosure and illegal greenhouse gas emissions focus on holding companies accountable for their environmental impact. These cases may involve allegations of misleading or incomplete disclosure of climate-related risks and emissions, and direct challenges to unlawful emissions practices. By enforcing compliance with environmental regulations and promoting corporate transparency, these lawsuits play a crucial role in advancing climate mitigation goals. Climate change mitigation is closely related to the climate litigation system. Implementing mitigation measures is crucial to achieving the "dual carbon" goals. However, from a long-term perspective, looking at climate issues on a scale of decades or even centuries reveals that excessive mitigation measures may not always be beneficial. Over-intervention could lead to further ecological disruption and secondary damage to the environment. Therefore, in designing a climate litigation system suited to China's national conditions, it is essential to consider not only the mitigation dimension but also the adaptation dimension. This balanced approach ensures that efforts to reduce emissions do not inadvertently harm ecosystems or undermine resilience to climate impacts. A well-designed climate litigation system should integrate both mitigation and adaptation strategies, providing a comprehensive legal framework to address the multifaceted challenges of climate change. This approach will help ensure that climate policies are effective, equitable, and sustainable, ultimately contributing to the long-term health and stability of the environment [38].

*3.2. Adaptation Dimension*

Climate change adaptation refers to enhancing the ability to cope with already occurring climate changes, minimizing social, economic, and life losses caused by climate change. Key measures include developing drought- and pest-resistant crops, controlling desertification, and creating climate insurance financial products [39]. These measures span various sectors such as agriculture, water supply, environmental protection, and government governance [40]. A strong legal framework underpins these fields, yet previous research has often focused solely on carbon reduction, lacking discussion on adaptation [41]. Therefore, the climate litigation system must not overlook the important dimension of adaptation. China has initially established a climate change policy framework, which extends from comprehensive departments to specialized ones in a top-down approach. At the national level, several policy documents, such as the "Action Plan for Carbon Peaking Before 2030" and the "National Plan for Addressing Climate Change", have set out the top-level design of the national adaptation policy framework. At the local level, government departments have formulated and implemented climate change adaptation plans based on these policy documents, while some provinces have also started drafting and promoting local regulations. Adaptation-focused climate litigation must emphasize coordination and alignment with ecological damage compensation lawsuits. Judicial interpretations and guiding cases, both existing and forthcoming, should provide courts with rules for handling adaptation-focused climate litigation [42]. The characteristics of climate change align most closely with the ecological damage compensation system. Environmental interests are primary concerns, and there is significant scientific uncertainty.

Legal liability is central to adaptation-focused climate litigation. This issue is complex because climate change impacts are often cross-temporal and cross-spatial, making it difficult to identify responsible parties and actions [43]. The urgency of clarifying government and corporate responsibilities, and protecting victims' rights, is especially pronounced during climate disasters. For governments, responsibility lies in formulating and implementing effective climate adaptation policies to minimize the impact of climate disasters. For corporations, adaptation-focused climate litigation necessitates considering the long-term effects of climate change in their operations and implementing appropriate adaptation measures. Protecting victims' rights is also crucial, as they often suffer significant loss of life and property during climate-related disasters [44]. Ensuring that victims can obtain appropriate compensation through legal channels involves not only establishing damage compensation mechanisms but also providing legal assistance and guaranteeing their right to participate in litigation. Thus, when courts hear adaptation-focused climate litigation cases, they must consider multiple factors, including policy design, risk assessment, liability determination, and rights protection [45]. Governments must establish clear policies and standards to guide corporations and all sectors of society in participating in climate adaptation actions. At the same time, the legal system should provide sufficient space for victims to protect their rights effectively. Only by doing so can the adaptation-focused climate litigation system effectively address climate change, particularly climate adaptation, minimizing personal injury and property damage caused by climate change and preventing the emergence of large-scale climate refugees.

## 4. Integration: Pathways to Constructing the Climate Litigation System

After thoroughly analyzing the jurisprudential foundation of the climate litigation system and examining it from the mitigation and adaptation dimensions, the next step is to explore how to integrate the unified framework to construct a climate litigation system that meets international standards. This system-building process requires establishing a preventive relief system and a multi-source legal framework at the substantive law level. At the procedural law level, procedures must be designed to be tailored to both mitigation and adaptation dimensions. Through the complementarity of substantive and procedural law, the preventive principle can be fully implemented, forming a climate litigation system that both mitigates and adapts to climate change. Establishing a preventive relief system is crucial for addressing the anticipatory nature of climate risks. This involves creating legal mechanisms for early intervention before significant harm occurs. Such mechanisms can include injunctions, preemptive litigation, and mandatory risk assessments. These tools ensure potential climate threats are identified and mitigated in advance, thereby preventing long-term environmental damage. The preventive approach is aligned with the precautionary principle, which advocates for proactive action in the face of scientific uncertainty regarding environmental harms. Building a multi-source legal framework at the substantive law level involves integrating domestic, foreign-related, and international laws. This framework must be robust enough to address the unique challenges climate change poses. Domestic laws must be aligned with international commitments, such as those outlined in the Paris Agreement, to ensure a coherent approach to climate governance. Foreign-related laws are essential for addressing cross-border environmental impacts and facilitating international cooperation on climate issues. International laws provide overarching principles and standards that guide national legislation and ensure global consistency in climate action. At the procedural law level, it is essential to design specific procedures tailored to both mitigation and adaptation efforts. For mitigation-focused litigation, procedures should facilitate the enforcement of emission reduction targets and compliance with environmental regulations. This can include streamlined processes for evidence collection, expert testimony, and scientific validation of climate data. For adaptation-focused litigation, procedures need to support the implementation of climate resilience measures, such as infrastructure improvements and disaster response planning. Legal processes should enable quick and effective responses to climate-induced emergencies, ensuring that affected communities receive timely support and compensation. By comprehensively constructing substantive and procedural laws, countries can create climate litigation systems that align with their national conditions while maintaining an international perspective. This system will coordinate domestic, foreign-related, and international law in climate governance, offering a global solution to climate litigation. Such a unified approach ensures that climate litigation effectively both prevents and addresses climate change impacts, thereby contributing to global sustainability and environmental justice. The development of this integrated climate litigation system can serve as a model for other nations, fostering international collaboration and strengthening global efforts to combat climate change.

### 4.1. Substantive Construction of Climate Litigation

In the future, countries need to innovate in the substantive rules of climate litigation to lay a solid foundation for procedural rules. Given climate change's global and complex nature, countries should establish a preventive relief system and a multi-source legal framework for climate litigation, implementing the precautionary principle and the principle of causal presumption. This approach would guide courts in applying new evidence rules, enabling plaintiffs to present their claims more effectively. By adopting these principles, legal systems can better address the uncertainties and risks associated with climate change, ensuring that preventative measures are taken to mitigate potential damages before they occur. Furthermore, a multi-source legal framework would integrate domestic laws with international climate commitments, providing a robust basis for enforcing climate policies and holding parties accountable for environmental harm. This comprehensive approach to climate litigation is essential for fostering international cooperation and ensuring that legal systems are equipped to handle the unique challenges posed by climate change. Such innovations are crucial for enhancing the effectiveness and fairness of climate litigation globally.

4.1.1. Establishing a Preventive Relief System

In a risk society—a concept referring to a society increasingly preoccupied with managing risks, especially those generated by modernization—the climate litigation relief system aims to correct structural flaws in climate risk governance, which currently cannot comprehensively address uncertain climate-related risks. Climate litigation's proactive nature can easily blur the boundaries between judicial power and other societal roles or responsibilities, rather than being framed as a right in itself. Therefore, the precautionary principle must be clearly defined in the substantive construction of climate litigation, forming a preventive judicial mechanism. This principle mandates that action be taken to prevent harm even if some cause-and-effect relationships are not scientifically established. For climate change, it is crucial to delineate the boundaries of preventive judicial relief, clarify the scope and authority compared to other relief methods, further implement the proactive judicial concept, strengthen the preventive value of the judiciary, solidify legal safeguards for judicial application in climate governance, and build a collaborative judicial system to help achieve climate goals on schedule. The key to a preventive relief system is to take measures in advance to prevent damage caused by climate change. Drawing on relevant experience from preventive environmental public interest litigation, it is necessary to develop a layered conceptual framework for climate preventive liability. This framework includes one core system (injunction system), two development stages (pre-litigation and during litigation), three different dimensions (theoretical, external, and substantive), four basic concepts (danger, risk, prevention, and precaution), and five forms of expression, meaning specific judicial remedies or actions, including injunction, preservation, cessation of infringement, removal of obstruction, and elimination of danger. The injunction system is central, serving as a legal tool to halt potentially harmful activities before significant damage occurs. The stages of pre-litigation and during litigation, referred to as "development stages," establish a framework that allows preventive measures to be implemented promptly at various points in the legal process, ensuring timely responses to emerging risks. The theoretical level involves foundational legal doctrines and principles, providing the conceptual basis for climate litigation; the external level encompasses policies and regulations beyond the judiciary, which influence and support judicial actions; and the substantive level focuses on the application of these theories and policies in specific legal cases and judgments, bridging theory and practice. The four basic concepts—danger, risk, prevention, and precaution—guide the judiciary in assessing the potential impacts of climate actions and deciding when to intervene. The five forms of expression of preventive liability—injunction, preservation, cessation of infringement, removal of obstruction, and elimination of danger—offer various legal remedies to prevent harm and ensure environmental protection. Ultimately, through this layered framework, the connotation and extension of climate preventive liability can be clarified, allowing for a more effective and responsive legal system to address the complexities of climate change.

4.1.2. Building a Multi-Source Legal Framework

In constructing a climate litigation system that meets international standards, a multi-source legal framework integrating both domestic and international law should be gradually developed, with an emphasis on establishing a proactive judicial system that effectively combines prevention and relief. A multi-source framework will help build a comprehensive and multidimensional climate litigation system better suited to the complexities of climate change in a globalized context. Therefore, the climate litigation system must not only consider domestic legal norms but also translate international climate change law and align it with domestic legislation. Currently, international climate negotiations are stalling, and some countries are eager to achieve their climate policy goals through judicial means. This creates external pressure for developing international climate change litigation [46]. Meanwhile, national climate goals intrinsically motivate developing international climate change litigation. Countries should develop their climate litigation systems along three paths: public law, private law, and cross-sectoral collaboration. A close analysis of climate change law reveals that it already has international legal characteristics. The current international legal framework is incomplete, and many countries and regions have not enacted specialized climate change laws. Therefore, due to the different stages of development, climate litigation can attempt to draw on various combinations of legal sources. For example, fundamental laws and constitutional provisions can be applied in its early stages, as they carry greater authority and priority in addressing climate change issues. If the litigation model evolves into a narrower form of climate litigation, specialized climate change laws should be enacted to support substantive law more effectively. This approach will help overcome legal barriers in judicial practice. By building a multi-source legal framework for climate litigation, judicial technical expertise can be applied to climate change cases to seek policy alignment, legal interpretation, reasoning, and principles. Past approaches, namely the "responsive judiciary" and the "proactive judiciary", differed in their strategies: the responsive judiciary reacted to climate-related cases brought before it, often limited to interpreting existing laws, while the proactive judiciary sought to shape climate policy by clarifying or even expanding judicial policies. However, despite being guided by climate policy goals, both approaches failed to uphold the jurisprudential function of climate justice in a standardized and systematic manner [47]. In the future, countries can encourage judicial bodies to pursue proactive climate justice and gradually work toward establishing legal principles and categorizing climate change policy pathways. In cases of conflict or rule exhaustion, the judiciary can perform specific functions to enhance climate justice: norm interpretation allows judges to clarify ambiguous climate laws or regulations; value declaration enables the court to affirm the importance of climate-related values, guiding future legal interpretation; and norm selection involves choosing among overlapping or competing norms to apply the most suitable one for advancing climate justice. These functions help ensure that climate justice remains adaptable and robust even when existing rules fall short.

*4.2. Procedural Construction of Climate Litigation*

Building a climate litigation system requires a strong procedural framework in addition to substantive groundwork. Once the substantive legal basis for climate litigation is established, clearly defining the litigants, scope of trial, and effective remedies becomes crucial. Furthermore, specialized climate change litigation case categories and corresponding litigation rules must be refined [48]. Therefore, in constructing a climate litigation system, it is essential to establish a judicial procedural mechanism for mitigation and adaptation by defining the right to sue. The procedural framework should establish a dual-step relief pathway where power and rights interact. Developing climate administrative public interest litigation at the power level involves enabling government agencies or public authorities to initiate legal actions to enforce climate regulations, ensuring that public power is used to protect the environment and address climate issues. In contrast, climate civil public interest litigation at the rights level empowers individuals or groups to file lawsuits to protect public environmental rights. This approach enables citizens to actively advocate for climate justice and hold polluters accountable. Together, these levels provide a comprehensive framework that leverages both governmental authority and individual rights to address climate-related concerns. By establishing a judicial procedural guarantee mechanism, the system would protect both public climate interests and victims' rights.

4.2.1. Procedural Construction of the Mitigation Dimension

As the legal maxim goes, "Where there is a right, there is a remedy; where there is no remedy, there is no right," emphasizing that rights must be enforceable through legal remedies to be meaningful. Therefore, a legal system without judicial remedies is inherently incomplete. In the context of judicial practice, ensuring enforceable climate rights is essential for achieving climate goals and effectively addressing climate change mitigation, as these rights provide the legal basis for holding entities accountable and enforcing climate-related obligations [49]. Therefore, the response to mitigation-focused climate litigation should center on rights-based remedies. The procedural framework should be led by both power and rights, requiring separate discussions for different relief pathways.

First, establish the relief pathway at the power level by developing climate administrative public interest litigation, which involves the use of state authority to initiate legal actions for the public good in climate-related cases. This approach should be implemented through five key perspectives: (1) In alignment with national conditions, climate administrative public interest litigation primarily adopts a dual-track model, incorporating both subjective litigation and objective litigation. Subjective litigation focuses on protecting specific individual or organizational rights that may be directly impacted by climate actions or policies, addressing concrete harm to particular parties. In contrast, objective litigation aims to uphold broader regulatory standards and public interests without a direct individual claim, focusing on ensuring that government actions and policies align with environmental laws and climate goals. This dual approach allows for a comprehensive legal response that addresses both individual rights and public regulatory compliance in climate-related cases. (2) Standing and Jurisdiction Scope: The scope of eligible defendants should be expanded beyond carbon emitters or direct polluting enterprises in individual cases. This is necessary for the construction of climate change litigation and helps protect broader social public interests while improving litigation effectiveness. (3) Jurisdiction Rules: In the administrative litigation jurisdiction system, the division of labor and authority for first-instance administrative cases is crucial. (4) Review Methods: Climate administrative public interest litigation should establish a preventive review mechanism that includes both legality and reasonableness reviews. Legality review examines whether government actions or corporate policies comply strictly with established climate laws and regulations, ensuring all decisions are within the legal framework. A reasonableness review, on the other hand, examines whether these actions are fair, suitable, and balanced in addressing climate issues, even if they technically meet legal standards. By combining these two types of reviews, this mechanism aims to ensure that climate-related decisions are both legally sound and aligned with broader principles of fairness and effectiveness in climate governance. (5) Judgment Forms: Based on the review methods described above, climate administrative litigation may result in decisions such as annulment judgments, enforcement judgments, judicial recommendations, administrative compensation, or damages. By constructing these categorized relief pathways at the power level, the administrative litigation pathway is designed to protect public climate interests, providing judicial procedural guarantees for the state to fulfill its public law obligations to mitigate climate change. This approach not only encourages governments and corporations to adopt more effective mitigation measures but also raises public awareness and engagement in climate issues, thereby strengthening the rule of law in climate governance.

Next, analyze and construct the rights-based relief pathway through climate civil public interest litigation, focusing on establishing exclusive jurisdiction within designated courts for specific types of climate civil public interest cases. This exclusive jurisdiction ensures that particular courts, equipped with specialized expertise in climate issues, handle these cases, allowing for consistent and informed rulings that uphold public environmental rights and effectively address climate-related harms. We suggest that the current IPCC carbon budget calculations are too lenient. It recommends adopting a stricter global carbon budget to ensure a higher probability of achieving the 1.5 °C target and highlights the potential challenges associated with many negative emission technologies [50]. The response should focus on separating litigable emission behavior, expanding the range of eligible defendants, and establishing comprehensive standards for determining the behaviour's legality under litigation [51]. Clear standards are crucial for determining the legality of the behavior under litigation in climate civil public interest litigation. These aspects offer climate civil litigation pathways to protect social public interests and victims' rights, providing a means to pursue liability for greenhouse gas emissions violations.

4.2.2. Procedural Construction of the Adaptation Dimension

In building the procedural framework for climate change adaptation, countries should strengthen their capacity in multiple areas through the development of long-term adaptation measures, plans, policies, projects, and programs [52]. The German Constitutional Court's ruling on climate protection addresses the intertemporal rights and obligations of freedom and climate policy, emphasizing the application of the precautionary principle within human rights. The study suggests that the German ruling indirectly impacts EU and international law [53]. Enabling them to better adapt to climate change and effectively handle adaptive climate litigation cases, minimizing the impact and losses of climate change on individuals, public health, and property. Adaptive climate litigation involves various types of cases. As climate change litigation is currently in its growth phase, incorporating more "peripheral" lawsuits and learning from their experience will be beneficial for the future development of specialized and systematic adjudication rules for climate change litigation [54]. For instance, in Leghari v. Federation of Pakistan, the court examined climate adaptation measures. It established an interdepartmental Climate Change Commission to ensure effective implementation, providing valuable insights for developing countries in judicial responses to climate change. Climate adaptation emphasizes improving the capacity to adapt to current and future climate change impacts. It involves a multi-stakeholder and multi-sectoral strategy, including technology, infrastructure, human resources, disaster response, and more, going beyond environmental justice. Environmental justice may apply to climate mitigation issues, but climate adaptation requires climate justice [55]. Adaptive climate litigation should focus on establishing a climate change impact assessment system, improving emergency response and rescue mechanisms, establishing a legal framework for post-disaster reconstruction and compensation, and establishing a climate damage relief mechanism.

5. Conclusions

In the context of China's active participation in global climate governance, judicial oversight of climate change has become a crucial component of both national and international climate strategies. To move beyond the limitations of past climate litigation research, it is essential to transcend the traditional paradigms of ecological damage compensation litigation and environmental public interest litigation. Establishing a third pathway for climate litigation in China involves integrating local practices with international experiences and fostering an East-West synergy. This approach not only leverages the rich experiences of climate litigation in the United States and other jurisdictions but also builds on China's existing practices, aiming to typologically analyze and construct the substantive and procedural mechanisms of climate litigation within the Chinese context. The proposed framework is based on a unified jurisprudential foundation of rights and obligations, considering climate mitigation and adaptation as its two primary dimensions. To establish an independent and effective climate litigation system, it is essential to analyze the necessity of such a system and compare it with existing environmental litigation structures. Traditional environmental litigation in China has primarily focused on compensating for ecological damage and pursuing environmental public interest litigation. However, these approaches often fail to address the complex, multifaceted nature of climate change and the urgent need for both preventive and adaptive measures. By constructing climate administrative and civil public interest litigation pathways and coordinating with ecological damage compensation litigation, the proposed framework aims to establish a Chinese-style climate litigation system that aligns with international standards and addresses global concerns. Developing specific procedural and substantive laws tailored to the unique challenges posed by climate change. For instance, administrative litigation can be used to hold government agencies accountable for failing to implement or enforce climate policies, while civil public interest litigation can target corporations and other entities responsible for significant greenhouse gas emissions. Localizing institutional arrangements for climate litigation is crucial for adapting international legal principles to China's legal and environmental context. This process includes integrating domestic laws with international commitments, such as those under the Paris Agreement, to ensure a cohesive and comprehensive legal framework. Additionally, developing new legal mechanisms and principles that reflect China's unique socio-economic and environmental conditions. This localized approach enhances not only the relevance and effectiveness of climate litigation in China but also contributes to the global discourse on climate justice by offering a distinctive model. Furthermore, Hong Kong's proactive climate litigation practices and robust ESG (Environmental, Social, and Governance) frameworks provide valuable insights that can enhance the effectiveness and international alignment of China's climate litigation system. Hong Kong's experience with mandatory ESG reporting and its strategic climate action plans offer practical examples of how legal and regulatory frameworks can be designed to promote sustainability and accountability. These practices highlight the importance of transparency, stakeholder engagement, and comprehensive reporting in achieving climate goals. By incorporating these elements, China's climate litigation system can ensure that both public and private entities are held accountable for their environmental impact. The proposed system aims to bridge the gap between theory and practice, offering a comprehensive legal response to climate change that can serve as a model for other nations. This involves developing robust legal frameworks and ensuring effective implementation and enforcement. For example, establishing specialized climate courts or tribunals could enhance the capacity of the judiciary to handle complex climate litigation cases. Additionally, training judges and legal practitioners in climate science and international environmental law can improve the quality and consistency of judicial decisions. Moreover, the system emphasizes the importance of a multi-source legal framework that integrates domestic and international laws, focusing on proactive judicial measures that combine prevention and relief. This includes creating legal mechanisms for early intervention, such as injunctions and preemptive litigation, to prevent significant environmental harm before it occurs. It also involves comprehensive compensation and adaptation mechanisms to address the impacts of climate change on vulnerable communities. In conclusion, by integrating local and international practices, the proposed climate litigation system aims to provide a comprehensive and effective legal framework for addressing climate change in China. This system aligns with international standards and addresses the specific challenges and opportunities within China's legal and environmental context. By fostering East-West synergy and drawing on the rich experiences of other jurisdictions, China can develop a robust and innovative approach to climate litigation that contributes to global climate governance and serves as a model for other nations.

**Acknowledgments**

I would like to express my sincere gratitude to School of Law of Fudan University for their invaluable administrative and technical support in the preparation of this paper. Special thanks to Sabin Center for Climate Change Law(Columbia University, USA) for providing the data used in Table 1, which significantly contributed to the analysis and findings of this study. I also appreciate the assistance of Law School of Zhongnan University of Economics and Law for their guidance and resources during the research process.

**Ethics Statement**

Not applicable.

**Informed Consent Statement**

Not applicable.

**Funding**

This research received no external funding.

**Declaration of Competing Interest**

The authors declare that they have no known competing financial interests or personal relationships that could have appeared to influence the work reported in this paper.

**References**

1. Peel J, Osofsky HM. *Climate Change Litigation: Regulatory Pathways to Cleaner Energy*; Cambridge University Press: Cambridge, UK, 2015; Volume 108, 301p.
2. McCormick S, Glicksman RL, Simmens SJ, Paddock L, Kim D, Whited B. Strategies in and outcomes of climate change litigation in the United States. *Nat. Clim. Chang.* **2018**, *8*, 829–833.
3. Loetscher A. Taking Carbon Culture to Court: Civil Lawsuits as Political Manifestoes in US Climate Change Litigation. *SPELL Swiss Pap. Engl. Lang. Lit.* **2019**, *38*, 43–59.
4. Hsu SL. A realistic evaluation of climate change litigation through the lens of a hypothetical lawsuit. *U. Colo. L. Rev.* **2008**, *79*, 701.
5. Myers B, Broderick J, Smyth S. Charting an Uncertain Legal Climate: Article III Standing in Lawsuits to Combat Climate Change. *Envtl. L. Rep. News Anal.* **2015**, *45*, 10509.
6. Setzer J, Vanhala LC. Climate change litigation: A review of research on courts and litigants in climate governance. *Wiley Interdiscip.Rev. Clim. Chang.* **2019**, *10*, e580.
7. Vanhala LC, Colli CW. Climate Change Litigation: New Directions in Law and Regulation. *Annu. Rev. Law Soc. Sci.* **2018**, *14*, 139–155.
8. Fisher E, Scotford E, Barritt E. The Legally Disruptive Nature of Climate Change. *Mod. Law Rev.* **2020**, *80*, 173–201.
9. Hague District Court. Urgenda Foundation v. State of the Netherlands, The Hague, The Netherlands, 2015. Available online: https://climatecasechart.com/non-us-case/urgenda-foundation-v-kingdom-of-the-netherlands/ (accessed on 15 December 2023).
10. Lahore High Court Green Bench. Leghari v. Federation of Pakistan, Lahore, Pakistani, 2015. Available online: https://climatecasechart.com/non-us-case/ashgar-leghari-v-federation-of-pakistan/ (accessed on 15 December 2023).
11. Cambridge University Press. *Litigating Climate Change through International Law*; Cambridge University Press: Cambridge, UK, 2020; Volume 33, pp. 933–951.
12. *Greenpeace Southeast Asia v. Carbon Majors*; Philippine Commission on Human Rights: Quezon, Philippines, 2015.
13. *Ralph Lauren 57 v. Byron Shire Council*; New South Wales Supreme Court: Sydney, Australia, 2016.
14. British Institute of International and Comparative Law. *Climate Change Litigation: Global Perspectives*; BIICL: London, UK, 2023.
15. Hong Kong Environmental Protection Department. *Hong Kong Climate Change Report*; Environmental Protection Department: Hong Kong, China, 2015.
16. Hong Kong Stock Exchange (HKEX). *ESG Reporting Guide*; Hong Kong Stock Exchange: Hong Kong, China, 2023.
17. Blumm MC, Wood MC. No ordinary lawsuit: climate change, due process, and the public trust doctrine. *Am. UL Rev.* **2017**, *67*, 1.
18. Heidari N, Pearce JM. A review of greenhouse gas emission liabilities as the value of renewable energy for mitigating lawsuits for climate change related damages. *Renew. Sustain. Energy Rev.* **2016**, *55*, 899–908.
19. International Bar Association. *Climate Change Justice and Human Rights Task Force Report: Achieving Justice and Human Right in an Era of Climate Disruption*; International Bar Association: London, UK, 2014.
20. Lytton TD. Using Tort Litigation to Enhance Regulatory Policy Making: Evaluating Climate-Change Litigation of Lessons from Gun-Industry and Clergy-Sexual-Abuse Lawsuits. *Tex. L. Rev.* **2007**, *86*, 1837.
21. Sabin Center for Climate Change Law. Climate Change Litigation Databases. Available online: https://climatecasechart.com/us-climate-change-litigation/ (accessed on 15 December 2023).
22. Kaswan A. Climate change and environmental justice: Lessons from the California lawsuits. *San Diego J. Clim. Energy L.* **2013**, *5*, 1.
23. United Nations Environment Programme and Sabin Center for Climate Change Law. *The Status of Climate Change Litigation: A Global Review*; United Nations Environment Programme: Nairobi, Kenya; Sabin Center for Climate Change Law: New York, NY, USA, 2017.
24. Adler DP. *U.S. Climate Change Litigation in the Age of Trump: Year One*; Columbia Public Law Research Paper; Sabin Center for Climate Change Law, Columbia Law School: New York, NY, USA, 2018.
25. Parliamentary Assembly of the Council of Europe. *Committee Proposes Draft of a New Protocol to the European Convention on Human Rights on the Right to a Healthy Environment*; Parliamentary Assembly of the Council of Europe: Strasbourg, France, 2021.
26. Stuart-Smith RF, Otto FEL, Saad AI, Lisi G, Minnerop P, Lauta KC, et al. Filling the evidentiary gap in climate litigation. *Nat. Clim. Chang.* **2021**, *11*, 651–655.
27. Setzer J, Higham C. *Global Trends in Climate Change Litigation: 2022 Snapshot*; Grantham Research Institute on Climate Change and the Environment: London, UK, 2022.
28. Gao L, Su D. Shifts in the Relationship between Judicial and Administrative Powers: A Perspective on U.S. Climate Litigation. *J. Nanjing Tech Univ. (Soc. Sci. Ed.)* **2023**, *5*, 36.

29. Supreme People's Court of the People's Republic of China. *China Environmental Resources Adjudication (2019)*; People's Court Press: Beijing, China, 2020.
30. Ali v. Federation of Pakistan, 2016. Available online: https://climatecasechart.com/non-us-case/ali-v-federation-of-pakistan-2/ (accessed on 15 December 2023).
31. Jamieson D. Climate Change, Responsibility, and Justice. *Sci. Eng. Ethics* **2010**, *16*, 431–445.
32. Viglione G. Climate lawsuits break new ground to protect the planet. *Nature* **2020**, *579*, 184–185.
33. Peel J, Lin J. Transnational climate litigation: The contribution of the Global South. *Am. J. Int. Law* **2019**, *113*, 679–726.
34. Omuko LA. Applying the Precautionary Principle to Address the 'Proof Problem' in Climate Change Litigation. *Tilburg Law Rev.* **2016**, *21*, 52–71.
35. Parker L, Mestre J, Jodoin S, Wewerinke-Singh M. When the kids put climate change on trial: Youth-focused rights-based climate litigation around the world. *J. Hum. Rights Environ.* **2022**, *13*, 64–89.
36. Otto F. The Science That Supports Climate Lawsuits. *Nature* **2021**, *597*, 169.
37. Setzer J, Benjamin L. Climate Litigation in the Global South: Constraints and Innovations. *Transnatl. Environ. Law* **2020**, *9*, 77–101.
38. Zhang Z. Preliminary Exploration of the Framework System for China's Climate Change Response Law. *J. Nanjing Univ. (Philos. Humanit. Soc. Sci. Ed.)* **2010**, *47*, 37–43.
39. United Nations Environment Programme (UNEP) and Sabin Center for Climate Change Law. Global Climate Litigation Report: 2023 Status Review. Available online: https://scholarship.law.columbia.edu/sabin_climate_change/202/ (accessed on 15 December 2023).
40. Peel J, Osofsky HM. A rights turn in climate change litigation? *Transnatl. Environ. Law* **2018**, *7*, 37–67.
41. IPCC. Climate Change 2014: Synthesis Report. In *Contribution of Working Groups I, II, and III to the Fifth Assessment Report of the Intergovernmental Panel on Climate Change*; Cambridge University Press: Geneva, Switzerland, 2014.
42. Banda ML, Fulton S. Litigating climate change in national courts: recent trends and developments in global climate law. *Envtl. L. Rep. News Anal.* **2017**, *47*, 10121.
43. Wang F, Harindintwali JD, Wei K, Shan Y, Mi Z, Costello MJ, et al. Climate change: Strategies for mitigation and adaptation. *Innov.Geosci.* **2023**, *1*, 61–95.
44. Madhuri, Kumar, R. Linkages Between Climate Change Adaptation and Development. In *Encyclopedia of the UN Sustainable Development Goals*; Springer Nature: Cham, Switzerland, 2021; pp 693–706.
45. Dellinger M. See you in court: Around the world in eight climate change lawsuits. *Wm. Mary Envtl. L. Pol'y Rev.* **2017**, *42*, 525.
46. Rogers N. Climate change litigation and the awfulness of lawfulness. *Altern. Law J.* **2013**, *38*, 20–24.
47. Klein N. *This Changes Everything: Capitalism vs. The Climate;* Simon Schuster, New York, USA, 2014.
48. Faure MG, Peeters M. *Climate Change Liability;* Edward Elgar Publishing, Cheltenham, UK, 2011.
49. Brunnée J, Toope SJ. Environmental Security and Freshwater Resources: Ecosystem Regime Building. *Am. J. Int. Law* **1997**, *91*, 26–59.
50. Ekardt F, Bärenwaldt M, Heyl K. The Paris Target, Human Rights, and IPCC Weaknesses: Legal Arguments in Favour of Smaller Carbon Budgets. *Environments* **2022**, *9*, 112.
51. Kysar DA. What Climate Change Can Do About Tort Law. *Environ. Law* **2011**, *41*, 1–71.
52. Hunter D, Salzman J, Zaelke D. *International Environmental Law and Policy*; Foundation Press: Goleta, CA, USA, 2011.
53. Ekardt F, Bärenwaldt M. The German Climate Verdict, Human Rights, Paris Target, and EU Climate Law. *Sustainability* **2023**, *15*, 12993.
54. Wiener JB. Global Environmental Regulation: Instrument Choice in Legal Context. *Yale Law J.* **1999**, *108*, 677–800.
55. Knox JH. Climate Change and Human Rights Law. *Va. J. Int. Law* **2010**, *50*, 163–218.